\documentclass[notitlepage, twoside,english,nofootinbib,preprintnumbers,aps,11pt,showpacs]{revtex4-1} \usepackage[a4paper, margin=1.5in]{geometry} \usepackage{amssymb} \usepackage{commath}
\usepackage{amsmath}
\usepackage{url}
\usepackage{float}
 \usepackage[utf8]{inputenc} \usepackage[T1]{fontenc} \usepackage{lmodern}
\usepackage{braket}
\usepackage{graphicx}
\usepackage{pictexwd,dcpic}
\usepackage{braket}
\usepackage{atbegshi,picture, afterpage}
\usepackage{lipsum}
\newcommand{\be}{\begin{equation}}
\newcommand{\ee}{\end{equation}}
\newcommand{\ba}{\begin{align}}
\newcommand{\ea}{\end{align}}
\begin{document}
\title{(Anti-)evaporation of Schwarzschild-de Sitter black holes revisited}
\author{Maciej Kolanowski}
\email[]{mp.kolanowski@student.uw.edu.pl}
\affiliation{\vspace{6pt} Institute of Theoretical Physics, Faculty of
Physics, University of Warsaw, Pasteura 5, 02-093 Warsaw, Poland}
  \date{\today}
\begin{abstract}
   {It is widely believed that in the presence of a positive cosmological constant, heavy black holes can exhibit non-standard behaviour, namely there is a possibility that such objects would grow instead of evaporating. We point out that all those results (obtained in different frameworks) rely heavily upon the identification of the Nariai spacetime with the Schwarzschild--de Sitter (Kottler) black hole. In this note we argue that it is an incorrect assumption. As a result, previous treatments need revisiting. In particular, we show that within effective action approach, there is no solution corresponding to the Schwarzschild--de Sitter black hole.}
\end{abstract}
\pacs{04.70.Dy, 04.50.Kd}
\maketitle
\section{Introduction}
It was proven in 1997 by Bousso and Hawking that some perturbations of Nariai spacetime can lead to anti-evaporation mechanism, growing of the horizon to be more precise \cite{Bousso:1997wi}. They considered an one--loop effective action obtained for S waves and in the limit of the large number $N$ of massless scalar fields. Similar scenarios were considered for two-dimensional black holes in dilaton gravity. It was also noticed that for primordial black holes (described by no boundaries condition in euclidean gravity) such effect would not be observed due to other, unstable perturbations. Later on, Nojiri and Odintsov pointed out that restriction to spherically symmetric fluctuations in integrating out was unjustified. They further investigated anti-evaporation of near Nariai spacetime using more general effective theories \cite{Nojiri:1998ue, Nojiri:1998ph, Nojiri:2000ja}. From the astrophysical point of view, the most interesting difference was that in the full 4-dimensional theory, the no boundary conditions were found to be consistent with the anti-evaporation, thus suggesting that primordial black holes could live much longer than previously expected. More recently, they also studied the same problem in the $f(R)$ framework \cite{Nojiri:2013su, Addazi:2017puj}. Their work was further generalized in \cite{Singh:2017qur, Fang:2018rte}. Similar scenario in two-dimensional dilaton gravity was considered in \cite{Buric:2000cj}. \\
In this paper we would like to point out a little confusion concerning quantum theory in the background of the Kottler spacetime\footnote{within this work we are going to treat names Kottler and Schwarzschild--de Sitter synonymously}, in particular the question of stability sketched above. Namely, extremal Schwarzschild--de Sitter black hole and Nariai solution are two distinct spacetimes. Their global properties are entirely different and only the second one seems to have any connection with astrophysical situations. This distinction was already made in \cite{Bousso:2002fq} but it was not followed by the comparative study of stability in both situations. Our aim is to fill this gap. We restrict our discussion to the approach \cite{Bousso:1997wi} due to its simplicity. It should be rather seen as a proof of concept than the final word on the topic. \newline
The rest of the paper is organized as follows: in the Sec. \ref{ext} we introduce all metrics relevant for our discussion and discuss their global structure. Then, in the Sec. \ref{bhee} we remind the reader effective equations derived in \cite{Bousso:1997wi} and investigate how extremal black hole shall look in this framework. We conclude our results in the Sec. \ref{sum}. 
\section{Considered spacetimes} \label{ext}
\subsection{Schwarzschild--de Sitter black hole}
All spherically symmetric solution to Einstein equations
\begin{equation}
    R_{\mu \nu} - \frac{1}{2}R g_{\mu \nu} = \Lambda g_{\mu \nu} \label{ee}
\end{equation}
were found by Kottler in 1918 \cite{Kottler1918}. Their form is surprisingly simple:
\begin{equation}
ds^2 = - \left(1 - \frac{2M}{r} - \frac{r^2}{\alpha^2} \right) d\tau^2  + \frac{dr^2}{1 - \frac{2M}{r} - \frac{r^2}{\alpha^2}} + r^2 d \Omega^2, \label{SdS}
\end{equation}
where $\alpha^2 = \frac{3}{\Lambda}$, $d \Omega^2$ is a standard metric on a $2$-sphere and $M$ is a parameter which can be identified as a mass of the black hole in question. As it is in the case of $\Lambda = 0$, from spherical symmetry automatically follows the existence of an additional Killing vector -- $\partial_\tau$. From now on, we are going to assume $\Lambda, M > 0$. \\
One can easily notice that spacetime described by \eqref{SdS} can have either two, one or none Killing horizons, depending on the relation between $M$ and $\Lambda$. For us, the most important case is so-called extremal black hole in which two roots of $g_{tt}$ are becoming one. It corresponds to the condition $9M^2 \Lambda = 1 $. It is well-known that the limit of one parameter families of spacetimes can depend non-trivially on used coordinates \cite{Geroch:1969ca} and so we describe two such limits (both corresponding to $9M^2 \Lambda \rightarrow 1^-$) in the following subsections. 
\subsection{Nariai spacetime}
Following \cite{Ginsparg:1982rs}, we introduce small parameter $\epsilon$:
\begin{equation}
    3 \epsilon^2 = 1 - 9M^2 \Lambda
\end{equation}
and new coordinates $(\psi, \chi)$ such that:
\begin{align}
    \begin{split}
        \tau &= \frac{1}{\epsilon \sqrt{\Lambda}} \psi \\
        r &= \frac{1}{\sqrt{6}} \left(1 - \epsilon \cos \chi - \frac{\epsilon^2}{6} \right).
        \label{nar_coord}
    \end{split}
\end{align}
Let us notice that \eqref{nar_coord} is diffeomorphism only for $\epsilon > 0$. This is the reason why we can have different limits. The metric reads:
\begin{equation}
    ds^2 = - \frac{1}{\Lambda}\left(1 + \frac{2}{3}\epsilon \cos \chi \right) \sin^2 \chi d\psi^2 + \frac{1}{\Lambda} \left(1 - \frac{2}{3}\epsilon \cos \chi \right) d \chi^2 + \frac{1}{\Lambda} \left( 1 - 2 \epsilon \cos \chi \right) d\Omega^2 + O(\epsilon^2).
\end{equation}
    In particular, taking the limit $\epsilon \rightarrow 0$, we obtain
    \begin{equation}
        ds^2 = - \frac{1}{\Lambda} \sin^2 \chi d\psi^2 + \frac{1}{\Lambda} d\chi^2  + \frac{1}{\Lambda} d\Omega^2,
    \end{equation}
    which is well-known Nariai solution \cite{Nariai1959}. It is simply a product spacetime $dS^2 \times S^2$. It follows that it is singularity-free and does not admit de Sitter asymptotics (in any sense) \cite{Ashtekar:2014zfa} and thus one can suspect that it is not a correct metric to describe what astrophysicist would mean as a black hole. \newline
    One can easily understand the sudden change of global properties in physical terms. Indeed, coordinate transformation \eqref{nar_coord} for $\epsilon \rightarrow 0^+$ can be seen as getting closer and closer to the horizon and so, one is 'loosing track of' what is very far. It can be checked that Nariai solution is exactly Near Horizon Geometry of the Schwarzschild--de Sitter black hole \cite{Kunduri:2013ana}. It follows that it still satisfies Einstein equations \eqref{ee}. \\
    It would be of interest also to understand what happens to the horizons in this limit. For simplicity, let us assume $\alpha = 1$. We can embed in a standard manner $2$-dimensional de Sitter spacetime into $3$-dimensional Minkowski spacetime:
    \begin{align}
    \begin{split}
        x^0 &= \sin \chi \sinh \psi \\
        x^1 &= \cos \chi \\
        x^2 &= \sin \chi \cosh \psi.
    \end{split}
    \end{align}
    in which $\partial_\psi =x^2 \partial_0 + x^0 \partial_2 $ is a generator of Lorentz transformation along $y$ direction. Not only its length is zero at the surfaces $\sin \chi = 0$ but the whole vector vanishes and our coordinate system becomes ill-defined. Knowing that its extention is $dS^2$ it is clear that $(x^0, x^1, x^2) = (0,\pm 1, 0)$ corresponds to the bifurcation sphere of the Killing horizon $(x^2)^2 - (x^0)^2 = 0$.
\subsection{Extremal Kottler black hole}
Instead of introducing new coordinates, one can take the limit $9M^2 \Lambda \rightarrow 1^-$ directly in the expression \eqref{SdS}, obtaining again regular spacetime. Its global structure was investigated in \cite{Podolsky:1999ts}. Its {\it scri} $\mathcal{I}^-$ is $S^3$ so it is asymptotically de Sitter spacetime (in contrast to the non-extremal case, in which $\mathcal{I}^- \cong \mathbb{R} \times S^2$). There is also $r=0$ singularity in the future. Thus, this spacetime contains only a black hole. Obviously, by a simple change of time orientation, one can obtain also a white hole solution. One should notice that $\partial_\tau$ is spacelike everywhere but on the horizon $r=3M$.
\\It seems to us that all those global properties suggest this limit as a correct description of extremal Schwarzschild--de Sitter black hole, especially seen by an observer located far from the event horizon.
\section{Bousso--Hawking effective equations}\label{bhee}
In the \cite{Bousso:1997wi}, authors considered general relativity with the cosmological constant $\Lambda > 0$ coupled to $N$ massless scalar fields.
They derived effective one-loop equations of motions in the large $N$ and s-wave approximation for a spacetime of the form:
\begin{equation}
    ds^2 = e^{2\rho} \left(-dt^2 + dx^2 \right) + e^{-2\phi}d\Omega^2,
\end{equation}
where $\rho$ and $\phi$ were functions of $t$ and $x$. It was assumed that $x$ is a coordinate on $S^1$ with period $2\pi$ but it was not used in the derivation of \eqref{eom} so we omit this assumption here. Obtained equations were following:
\begin{align} \label{eom}
    &\left(1-\frac{\omega \kappa}{4}e^{2\phi}\right) \partial^2 \phi - 2\left(\partial \phi\right)^2 - \frac{\kappa}{4}e^{2\phi} \partial^2 Z - e^{2\rho +2\phi} \left(\Lambda e^{-2\phi} - 1 \right) = 0, \\
    &\left(1-\frac{\omega \kappa}{4}e^{2\phi}\right) \partial^2 \rho - \partial^2 \phi + (\partial \phi)^2 + \Lambda e^{2\rho} = 0 \\
    &\partial^2 Z - 2 \partial^2 \rho = 0, \label{Zzz}
\end{align}
together with two constraint equations:
\begin{align}
    &\left(1-\frac{\omega \kappa}{4}e^{2\phi}\right) \left(\delta^2 \phi - 2\delta \phi \delta \rho \right) - \left( \delta \phi\right)^2 = \frac{\kappa}{8}e^{2\phi}
    \left[ (\delta Z)^2 + 2 \delta^2 Z - 4\delta Z \delta \rho
    \right], \\
    &\left(1-\frac{\omega \kappa}{4}e^{2\phi}\right) \left(\dot{\phi}' - \dot{\rho} \phi' - \rho' \dot{\phi}  \right) - \dot{\phi}\phi' = \frac{\kappa}{8}e^{2\phi} \left[ \dot{Z} Z' + 2 \dot{Z}' - 2 \left(\dot{\rho} Z' + \rho' \dot{Z} \right)
    \right]. \label{const_Z}
\end{align}
The notation is as follows:
\begin{align}
    \begin{split}
        &\partial f \partial g = f'g' - \dot{f}\dot{g}, \ \  \partial^2 g = g'' - \ddot{g}. \\
        &\delta f \delta g =  f'g' + \dot{f}\dot{g}, \ \  \delta^2 g = g'' + \ddot{g},
    \end{split}
\end{align}
$' = \frac{\partial}{\partial x}$, $\dot{\ } = \frac{\partial }{\partial t}$, $Z$ is an auxiliary field introduced to make effective action local, $\kappa = \frac{2}{3}N$ and $\omega$ is a coupling constant for the term $\phi R$. \\
The strategy taken in previous work was to find generalized Nariai solution to the above equations (understood as a product spacetime $dS^2 \times S^2$ with some effective radii) and to analyze its stability. In this paper, however, we are more interested in a usual Schwarzschild--de Sitter so we are going to assume that
\begin{equation}
    e^{-2\phi} = r^2,
\end{equation}
where $r$ and $\rho$ are some functions of $t$ only. Coordinate $x$ would correspond to the $\tau$ coordinate in Sec. \ref{ext}. It allows for drastic simplification of equations (\ref{eom}-\ref{const_Z}). While we relegate technical calculations to the Appendix, let us formulate the main result: \newline
{\bf Theorem} {\it There is no solution to (\ref{eom}-\ref{const_Z}) with Killing vector $\partial_x$ and asymptotic behaviour:}
\begin{equation}
    ds^2 = -  \frac{dr^2}{B^2 r^2} + A^2 r^2 dx^2 + r^2 d\Omega^2
\end{equation}
{\it for large $2-sphere$ radius $r$ where $A,B$ are non-zero constants.} \\
This expansion is needed for the existence of conformal factor which would result in structure of {\it scri} such as Schwarzschild--de Sitter's \cite{Ashtekar:2014zfa}.
\subsection{Discussion}
Above theorem suggests that such an effective approach to black holes is not correct to describe backreaction due to Hawking radiation. Let us emphasize that we have demonstrated the lack of any Schwarzschild--de Sitter-like solutions, not only extremal ones. One could fear that this framework is ill-suited to addressing the question of stability of the black hole (in addition to concerns raised in \cite{Nojiri:1999br}). Perturbations considered previously in the literature look small near the horizon but they are in fact getting bigger far away. \newline
Let us empasize that our result should be treated only as a proof of principle that those two spacetimes are different thus can yield different behaviours. It would be of interest to check whether analyzed discrepancies persist without s-wave approximation and also in $f(R)$ theories. If the answer was positive, it  could lower expected lifetime of a primordial black hole. Furthermore, let us notice that for the extremal Kottler metric, coordinate $r$ plays rather a role of time than radius. Thus, interpreting any change of Schwarzschild radius $r_S \mapsto r_S + \epsilon$ under some perturbation as a growing of horizon seems not complete, it could also correspond to some time translation.
\section{Summary and future work} \label{sum}
In this paper, we have discussed various notions of extremal black holes present in the literature. We point out the differences between them, especially in the context of their asymptotics. From this we conclude that the extremal Schwarzschild--de Sitter black hole investigated in \cite{Podolsky:1999ts} is more physically relevant than Nariai solution. Unfortunately, the effective theory of Bousso and Hawking does not contain such a spacetime which suggests problem with the conclusions drawn from equations $($\ref{eom}-\ref{const_Z}$)$. In particular, anti-evaporation effect of near-extremal black holes can be in fact only a mathematical artifact. This analysis is still prelimnary. In particular, since $f(R)$ theories can have Schwarzschild--de Sitter solutions, the question of stability is much less trivial therein. Moreover, one would need not only address perturbation theory but also the physical predictions of particular model. We hope to investigate it further in the future.
\begin{acknowledgments}
Author would like to thank Wojciech Kamiński, Jerzy Lewandowski and Sergei Odintsov for frutiful disscusions. This work was financed from budgetary funds for science in 2018-2022 as a research project under the program "Diamentowy Grant".
\end{acknowledgments}
\appendix
\section{Calculations}
We are interested in \eqref{eom}. After replacing $\partial^2 Z$ by $2 \partial^2 \rho$ due to \eqref{Zzz} and assuming everything depends solely on $t$, we obtain the following equation:
\begin{equation}
   -\left(1-\frac{\omega \kappa}{4}e^{2\phi}\right) \ddot{\phi} - 2\dot{\phi}^2 + \frac{\kappa}{2}e^{2\phi} \ddot{\rho} - e^{2\rho +2\phi} \left(\Lambda e^{-2\phi} - 1 \right) = 0. \label{app_eq1}
\end{equation}
We introduce invertible function $r(t)$ such that $e^{-2\phi} = r^2$. For simplicity, let us denote $e^{\rho} = g(t)$ and $\frac{\partial r}{\partial t} = g(t) h(t)$. Changing coordinates from $t$ to $r$, the metric takes the form:
\begin{equation}
    ds^2 = - \frac{dr^2}{h^2} + g^2 dx^2 + r^2 d\Omega^2,
\end{equation}
which is well-suited for proving the main result of our paper. Now, writing \eqref{app_eq1} in terms of $r, g$ and $h$, one obtains:
\begin{align}
\begin{split}
    &- \left(1 - \frac{\omega \kappa}{4 r^2} \right) gh\left(\frac{g h}{r^2} - \frac{g_{,r} h + g h_{,r}}{r} \right) - \frac{2g^2 h^2}{r^2} \\
    &+ \frac{\kappa h g}{2r^2} \left(h_{r} g_{,r} + h g_{,rr} \right) - \frac{g^2}{r^2} \left( \Lambda r^2 -1 \right) = 0
    \end{split}
\end{align}
Assuming that for large $r$ $g = A r + O(r^{1-\epsilon})$, $h = B r + O(r^{1-\epsilon})$ we get:
\begin{equation}
    -A^2 (2B^2 + \Lambda) r^2 + O(r^{2 - \epsilon}) = 0.
\end{equation}
Since $\Lambda > 0$, we obtain contradiction and thus our main theorem is true. 
\bibliography{bibl.bib}
%\printbibliography
\end{document}